\documentclass[aps,prl,reprint,preprintnumbers,superscriptaddress,amsmath,amssymb,bibnotes,longbibliography]{revtex4-1}

\usepackage{graphicx}
\usepackage{dcolumn}
\usepackage{bm}
\usepackage[colorlinks,linkcolor=blue,anchorcolor=blue,citecolor=blue,urlcolor=blue,filecolor=blue,menucolor=blue,runcolor=blue]{hyperref}
\begin{document}

\title{ Evolution of charge density wave order and superconductivity under pressure in LaPt$_2$Si$_2$}

\author{B. Shen}
\affiliation  {Center for Correlated Matter and Department of Physics, Zhejiang University, Hangzhou 310058, China}

\author{F. Du}
\affiliation  {Center for Correlated Matter and Department of Physics, Zhejiang University, Hangzhou 310058, China}

\author{R. Li}
\affiliation  {Center for Correlated Matter and Department of Physics, Zhejiang University, Hangzhou 310058, China}

\author{A. Thamizhavel}
\affiliation  {DCMP $\&$ MS, Tata Institute of	Fundamental Research, Mumbai 400005, India}

\author{M. Smidman}
\affiliation  {Center for Correlated Matter and Department of Physics, Zhejiang University, Hangzhou 310058, China}

\author{Z. Y. Nie}
\affiliation  {Center for Correlated Matter and Department of Physics, Zhejiang University, Hangzhou 310058, China}

\author{S. S. Luo}
\affiliation  {Center for Correlated Matter and Department of Physics, Zhejiang University, Hangzhou 310058, China}

\author{T. Le}
\affiliation  {Center for Correlated Matter and Department of Physics, Zhejiang University, Hangzhou 310058, China}

\author{Z. Hossain}
\affiliation  {Department of Physics, Indian Institute of Technology, Kanpur 208016, India}

\author{H. Q. Yuan}
\email[Corresponding author: ]{hqyuan@zju.edu.cn}
\affiliation  {Center for Correlated Matter and Department of Physics, Zhejiang University, Hangzhou 310058, China}
\affiliation  {Collaborative Innovation Center of Advanced Microstructures, Nanjing University, Nanjing, 210093, China}

\date{\today}

\begin{abstract}
We report  measurements of the electrical resistivity and ac magnetic susceptibility of  single crystalline LaPt$_2$Si$_2$ under pressure, in order to investigate the interplay of superconductivity and CDW order. LaPt$_2$Si$_2$ exhibits a first order phase transition from a tetragonal to orthorhombic structure, accompanied by the onset of  CDW order below $T_{\rm{CDW}}$ = 76~K, while superconductivity occurs at a lower temperature of $T_{\rm{c}}$ = 1.87~K. We find that the application of pressure initially  suppresses the CDW transition, but enhances  $T_{\rm{c}}$. At pressures above 2.4~GPa, CDW order vanishes, while both  $T_{\rm{c}}$ and the resistivity $A$-coefficient reach a maximum value around this pressure. Our results suggest that the occurrence of a superconducting dome can be accounted for within the framework of BCS theory, where there is a maximum in the density of states upon the closure of the CDW gap. 

\end{abstract}

\maketitle

\section{\uppercase\expandafter{\romannumeral1}. INTRODUCTION}

 The continuous suppression of a phase transition to zero temperature, which is accompanied by superconductivity (SC) in some cases,  has long been one of the most intriguing and extensively studied phenomena in condensed matter physics. One recent example is the quantum criticality related to a charge density wave (CDW) phase \cite{NP_CuTS, Pressure_TS, NP2017_LPI}. CDW order corresponds to a condensate with periodic modulations of the electron density, often found in low-dimensional metallic systems. CDW phases have been well documented in various compounds,  and in many cases these systems also exhibit superconductivity either in ambient conditions or upon tuning with non-thermal parameters such as pressure \cite{RMPA15, COLEMAN1973, CDW1975, CDWCu, Review2002, Gabovich_2001}. Typically,  the superconductivity is found to be abruptly enhanced upon suppressing the  CDW order \cite{Gabovich_2001}. This can be well understood in the framework of Bardeen-Cooper-Schrieffer (BCS) theory, since CDW order gaps out certain regions of the Fermi surface, and as such  the suppression of CDW order  leads to an enhancement of the superconducting transition temperature $T_{\rm{c}}$ due to the sudden enhancement of the density of states at the Fermi level $N(E_F)$.  
 
 On the other hand, a different scenario for the interplay of CDW and SC  is the occurrence  of a superconducting dome on the edge of a CDW/structural instability, as  proposed for  $T$Se$_2$ ($T$ = Ti, and Ta) \cite{NP_CuTS, Pressure_TaS, Pressure_TS, Pressure_PdTaSe},  Lu{(Pt$_{1-x}$Pd$_x$)}$_2$In \cite{NP2017_LPI}, $o$-TaS$_3$ \cite{Pressure_TaS3} and {(Ca,Sr)}$_3$Ir$_4$Sn$_{13}$ \cite{Malte2012CIS}.  SC in conjunction with non-Fermi liquid behaviors   is often found in close proximity to a magnetically ordered phase in strongly correlated electronic systems, such as heavy fermion compounds, high $T_{\rm{c}}$ cuprates, and iron-based superconductors \cite{UncoSC_Stewart}. In addition, it has been proposed that the accumulation of  entropy near a quantum critical point (QCP), due to  quantum critical fluctuations, may give rise to novel phases such as superconductivity \cite{NP2017}. However, one difference between a magnetic QCP in strongly correlated electronic systems and that of a QCP in CDW compounds, is that non-Fermi liquid behaviors seem to be absent in the vicinity of a CDW QCP  \cite{NP_CuTS, Pressure_TS, NP2017_LPI}. Furthermore, the observation of CDW domain walls above the superconducting dome and the  separation of the CDW QCP and superconductivity in pressurized 1$T$-TiSe$_2$ seems to challenge the view that the appearance of a superconducting dome is associated with the CDW QCP \cite{NP_TS_Domain}. Therefore, it is still an open question as to whether critical quantum fluctuations can facilitate SC in the vicinity of a CDW QCP.

LaPt$_2$Si$_2$ belongs to the $M$$T_2$$X_2$ ($M$ = rare earth/alkaline earth, $T$ = transition metal, $X$ = Si or Ge) family crystallizing in the CaBe$_2$Ge$_2$-type structure. Several compounds with this  structure exhibit both CDW order and SC, such as SrPt$_2$As$_2$ \cite{Sr122} and BaPt$_2$As$_2$ \cite{jiang2014}. SrPt$_2$As$_2$ exhibits a CDW transition at around 470~K and becomes superconducting at 5.2~K \cite{Sr122}, while BaPt$_2$As$_2$ undergoes a first order structural transition at 275~K and a bulk superconducting transition at $T_{\rm{c}}$ = 1.33~K \cite{jiang2014}. Furthermore, BaPt$_2$As$_2$ exhibits a complex temperature-pressure phase diagram with multiple pressure-induced transitions at high temperature and abrupt changes in $T_{\rm{c}}$ which coincide with the high temperature phase transitions \cite{GuoBa}. 

Studies on polycrystalline  LaPt$_2$Si$_2$ show that it undergoes a structural phase transition upon cooling, from a tetragonal to orthorhombic phase accompanied by a CDW phase transition at around $T_{\rm{CDW}}$ = 112~K, and is superconducting below $T_{\rm{c}}$ = 1.8~K \cite{2013CDW}. Meanwhile in single crystals, the CDW order was found at a lower temperature of 80~K \cite{Gupta_2017}. Additional  superlattice reflections corresponding to the tripling of the unit cell along the [110] direction\cite{2013CDW} strongly suggests the presence of CDW order. Furthermore,  CDW order is detected in other physical quantities such as the thermopower and thermal conductivity \cite{2018TT}.   First principles calculations suggest that the Fermi surface of LaPt$_2$Si$_2$ is quasi two-dimensional, and that there is coexistence of CDW order and SC in the [Si2-Pt1-Si2] layer  \cite{2013FS, 2015FSSciRep}. The coexistence of the two orders and a partially opened CDW gap on the Fermi surface were confirmed by nuclear magnetic resonance (NMR) experiments\cite{2018NMR}. $\mu$SR measurements of LaPt$_2$Si$_2$ indicate that the superconductivity is well described by a two-gap $s$-wave model rather than a single isotropic gap \cite{2018uSR}. A pressure study of LaPt$_2$Si$_2$ shows the decrease of $T_{\rm{CDW}}$  and increase of $T_{\rm{c}}$ with pressure, but the maximum pressure applied was not high enough to fully suppress the CDW order\cite{gupta2019electrical}. It is therefore of great interest to apply higher pressure using a diamond anvil cell so as to investigate the interplay of CDW and SC in LaPt$_2$Si$_2$.    Here, we report  electrical transport and ac susceptibility measurements of single crystals of LaPt$_2$Si$_2$ under pressures up to 7~GPa, and we construct the temperature-pressure phase diagram.    
        
\section{\uppercase\expandafter{\romannumeral2}. EXPERIMENTAL METHODS}

Single-crystals of LaPt$_2$Si$_2$ were synthesized using the Czochralski method, as described in Ref.\cite{Gupta_2017}. The specific heat down to 0.4~K was measured in a Quantum Design Physical Property Measurement system (PPMS) with a $^3$He insert, using a standard pulse relaxation method. The resistivity and ac susceptibility measurements were performed in a Teslatron-PT system with an Oxford $^3$He refrigerator, with a temperature range of 0.3~K to 300~K and a maximum applied magnetic field of 8~T. Single crystals of LaPt$_2$Si$_2$ were polished and cut into rectangular pieces with  approximate dimensions 180~$\mu$m $\times$ 80~$\mu$m  $\times$ 30~$\mu$m, loaded into a BeCu diamond anvil cell (DAC) with a 800-$\mu$m-diameter culet. A  100-$\mu$m-thick preindented CuBe gasket was covered with Al$_2$O$_3$ for electrical insulation and a 400-$\mu$m-diameter hole was drilled as the sample chamber. Daphne oil 7373 was used as the pressure-transmitting medium. The DAC was loaded together with several small ruby balls for pressure determination using the ruby fluorescence method at room temperature. For electrical transport measurements, four 15~$\mu$m-diameter gold wires were glued to the samples with silver epoxy paste and the resistivity was measured using the standard four-probe method. For ac susceptibility measurements, a 3~Oe magnetic field was generated by the driven coil placed outside the DAC, and two counter wound pickup coils were used to pick up the magnetic signal, with the sample in the center of one of these coils.

\section{\uppercase\expandafter{\romannumeral3}. RESULTS}

\begin{figure}
	\includegraphics[angle=0,width=0.49\textwidth]{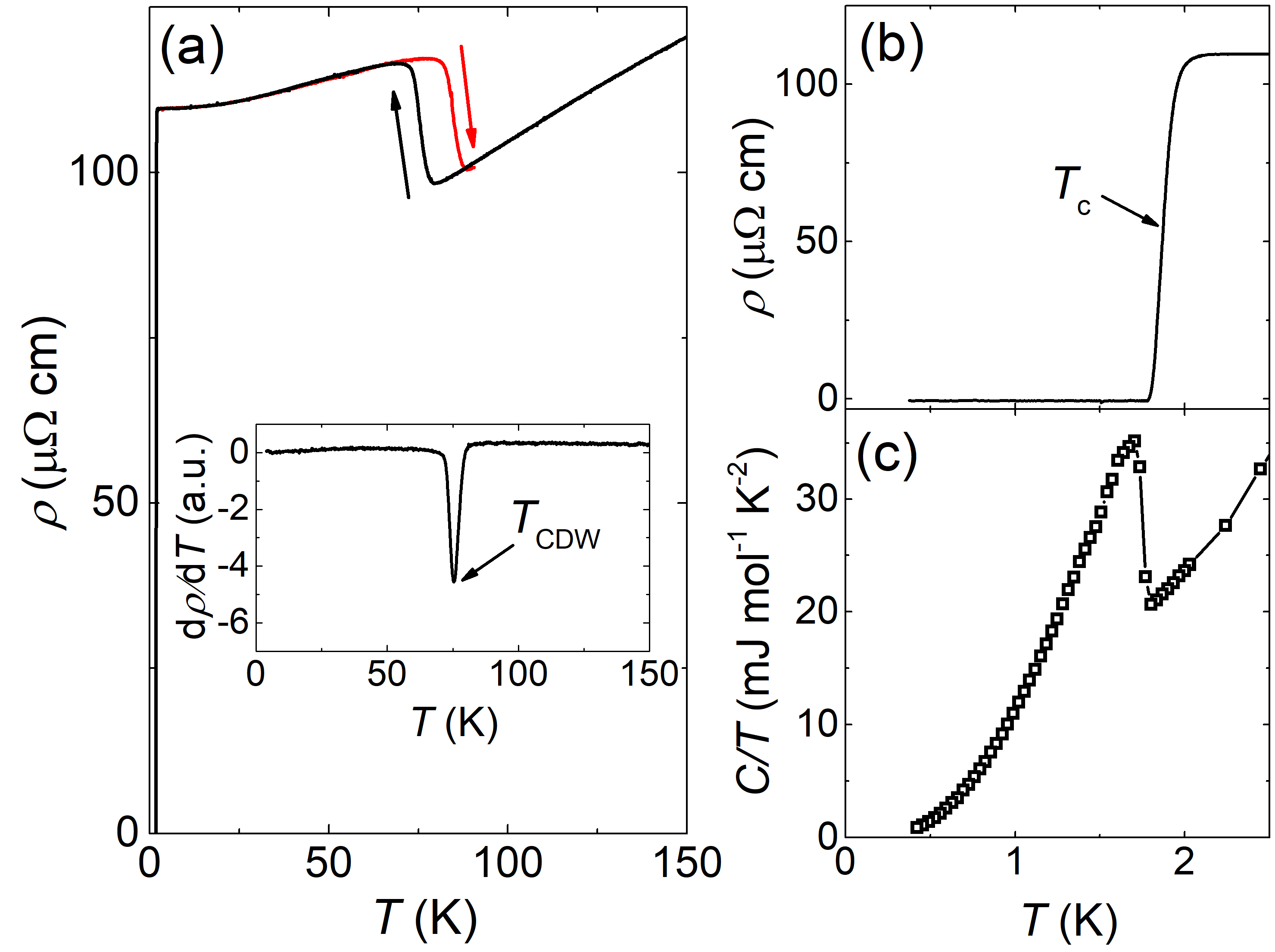}
	\vspace{-12pt} \caption{\label{Figure1}(color online)  (a) Resistivity at ambient pressure of LaPt$_2$Si$_2$ as a function of temperature from 150~K down to 0.3~K. The black (red) arrow denotes the data taken upon cooling (warming). (inset) d$\rho$/d$T$ on cooling which was used to define $T_{\rm{CDW}}$. Low temperature (b)  resistivity, and (c) specific heat as $C/T$, which both show the presence of a superconducting transition. The arrow in (b) marks the position of  $T_{\rm{c}}$, corresponding to the midpoint of the resistivity drop.}
	\vspace{-12pt}
\end{figure}

  Figure \ref{Figure1}(a) shows the temperature dependence of the resistivity $\rho(T)$ from 150~K down to 0.3~K at ambient pressure. The resistivity as a function of temperature shows a clear step-like first order phase transition at $T_{\rm{CDW}}$ = 76~K, with a hysteresis loop in $\rho(T)$ between measurements performed upon cooling down and warming up. Here $T_{\rm{CDW}}$ is defined as the minimum of d$\rho$/d$T$ on the cooling curve, as displayed in the inset of Fig. \ref{Figure1}(a). At low temperature, the compound undergoes a superconducting  transition at around $T_{\rm{c}}$ = 1.87~K, defined as the midpoint of the resistivity drop [Fig. \ref{Figure1}(b)]. This also corresponds to the transition observed in the heat capacity $C(T)/T$ [Fig. \ref{Figure1}(c)]. Moreover, the sharp nature of the transitions in $\rho(T)$ and $C(T)/T$ indicates a good sample quality. Note that the single crystals of  LaPt$_2$Si$_2$ studied here show a much lower $T_{\rm{CDW}}$ but higher $T_{\rm{c}}$ than the polycrystalline samples previously reported in Ref.\cite{2013CDW}. These differences are likely due to a slight variation in the lattice constants (pressure effect) or sample homogeneity/composition between the polycrystalline and single crystal samples, which might significantly change the values of $T_{\rm{CDW}}$ and $T_{\rm{c}}$, as shown below.

\begin{figure}
	\includegraphics[angle=0,width=0.49\textwidth]{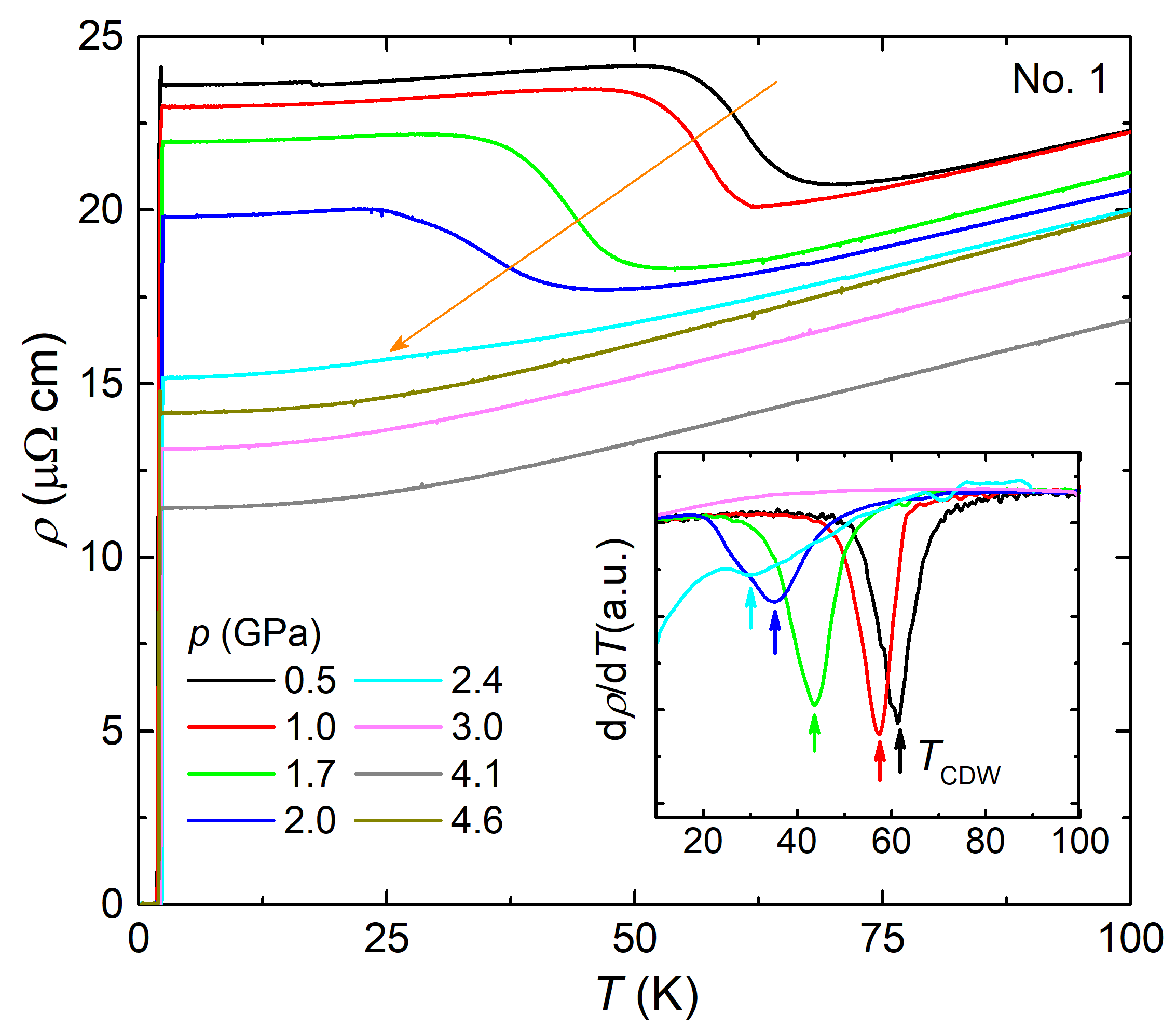}
	\vspace{-12pt} \caption{\label{Figure2}(color online)   Temperature dependence of the resistivity of LaPt$_2$Si$_2$ sample No. 1 under various pressures, measured upon  cooling.  The inset shows the resistivity derivative d$\rho$/d$T$, where the arrows mark the CDW transition temperatures.}
	\vspace{-12pt}
\end{figure}
  
  In order to track the evolution of CDW order, we performed resistivity and ac susceptibility measurements under pressure, where Fig. \ref{Figure2} displays the resistivity of sample No. 1 from 100~K to 1.6~K under various pressures. With the application of pressure, the CDW order is suppressed to lower temperature and the transition becomes broadened. At 2.4~GPa, the CDW transition is hardly seen in the resistivity data, but is still visible in the derivative of the resistivity d$\rho$/d$T$  as seen in the inset of Fig. \ref{Figure2}. With further increasing pressure, there is no signature of the CDW transition in both the resistivity and its derivative.  At 3~GPa, $\rho(T)$ shows metallic behavior down to the superconducting transition.

  \begin{figure}
  	\includegraphics[angle=0,width=0.49\textwidth]{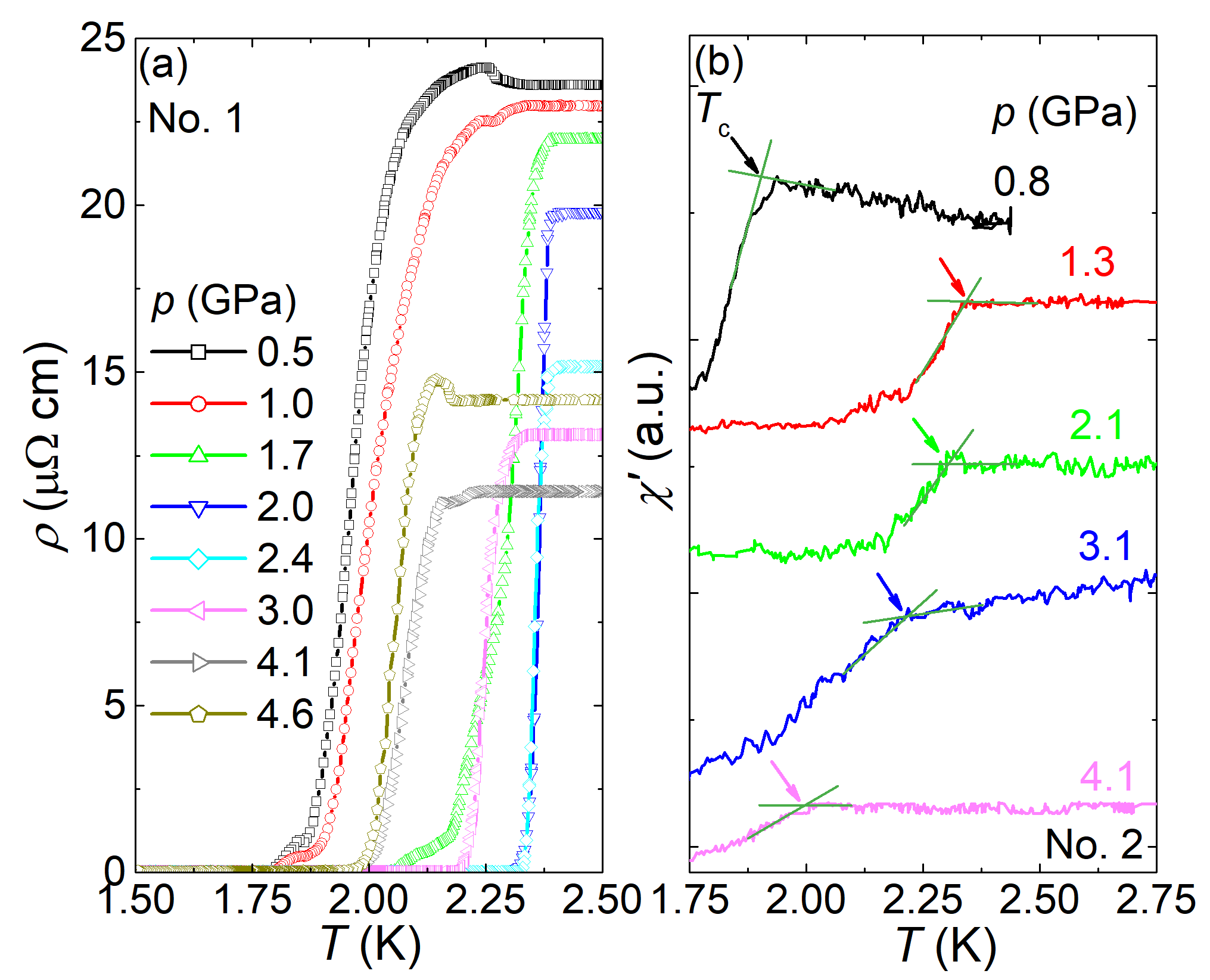}
  	\vspace{-12pt} \caption{\label{Figure3}(color online)   Evolution of the superconducting transition of LaPt$_2$Si$_2$ under various pressures. The low temperature part of  (a) $\rho(T)$ of sample No. 1 and (b) the real part of the ac susceptibility of sample No. 2 are displayed. The arrows in panel (b) indicate the position of  $T_{\rm{c}}$.}
  	\vspace{-12pt}
  \end{figure}

  \begin{figure}[t]
	\includegraphics[angle=0,width=0.49\textwidth]{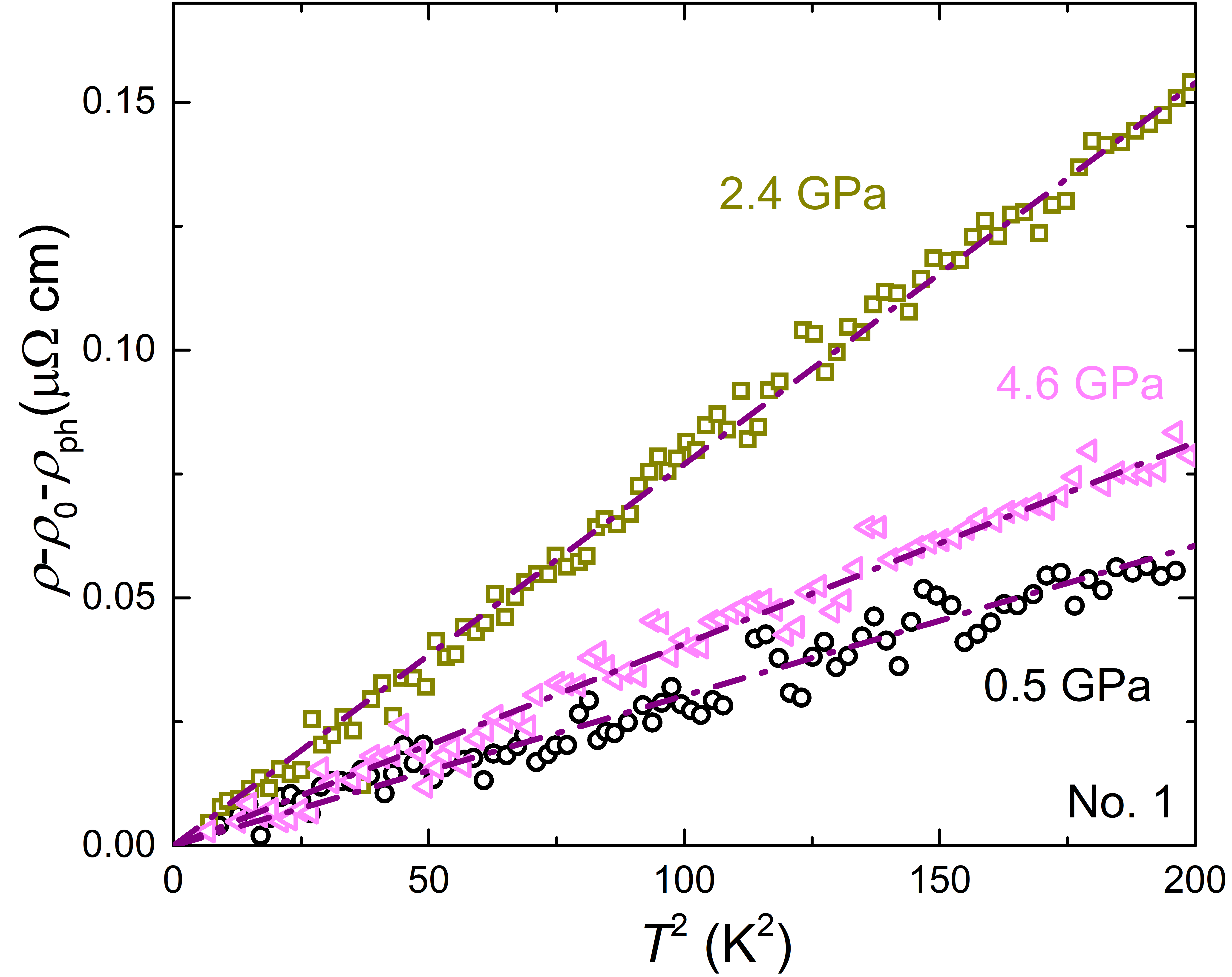}
	\vspace{-12pt} \caption{\label{Figure4}(color online)   Resistivity versus $T^2$ of sample No. 1 at several pressures, where  the residual resistivity $\rho_0$ and the phonon contribution $\rho_{\rm{ph}}$ are subtracted. The dashed-dotted lines show the results from fitting Fermi liquid behavior in the normal state.}
	\vspace{-12pt}
\end{figure}

 Figure \ref{Figure3}(a) displays the low temperature behavior of the resistivity of sample No. 1. Upon applying pressure, $T_{\rm{c}}$ initially increases to higher temperatures and reaches a maximum value of $T_{\rm{c}}^{\rm{max}}$ = 2.36~K at around 2.0~GPa and 2.4~GPa. The maximum $T_{\rm{c}}$ is close to the pressure region where the CDW transition disappears. We also performed ac susceptibility measurements on LaPt$_2$Si$_2$ [Fig. \ref{Figure3}(b)], from which $T_{\rm{c}}$($p$) exhibits similar behavior to the resistivity. Meanwhile, $\rho(T)$ in the normal state of LaPt$_2$Si$_2$ under pressure can be fitted using $\rho(T)$ = $\rho_0$ + $A$$T^2$ + $b$$T^5$, across a wide temperature range. Here, $\rho_0$ is the residual resistivity, $A$ is the resistivity coefficient related to the Fermi-liquid state, and the last term corresponds to the electron-phonon scattering $\rho_{\rm{ph}}$ at low temperature. The latter term is valid since the data were fitted at temperatures much lower than the Debye temperature ($\theta_D$ = 221.3~K) \cite{Gupta_2017}. The fitted values of $b$ and $A$ are on the order of ~10$^{-8}$$\mu\Omega$~cm~K$^{-5}$  and ~10$^{-4}$$\mu\Omega$~cm~K$^{-2}$ respectively, suggesting that electron-electron scattering dominates the resistivity. Figure \ref{Figure4} displays the resistivity after subtracting $\rho_0$ and $\rho_{\rm{ph}}$ as a function of $T^2$. The data exhibits a quadratic temperature dependence, indicating a Fermi liquid ground state at all pressures.
  
  \begin{figure}[t]
	\includegraphics[angle=0,width=0.49\textwidth]{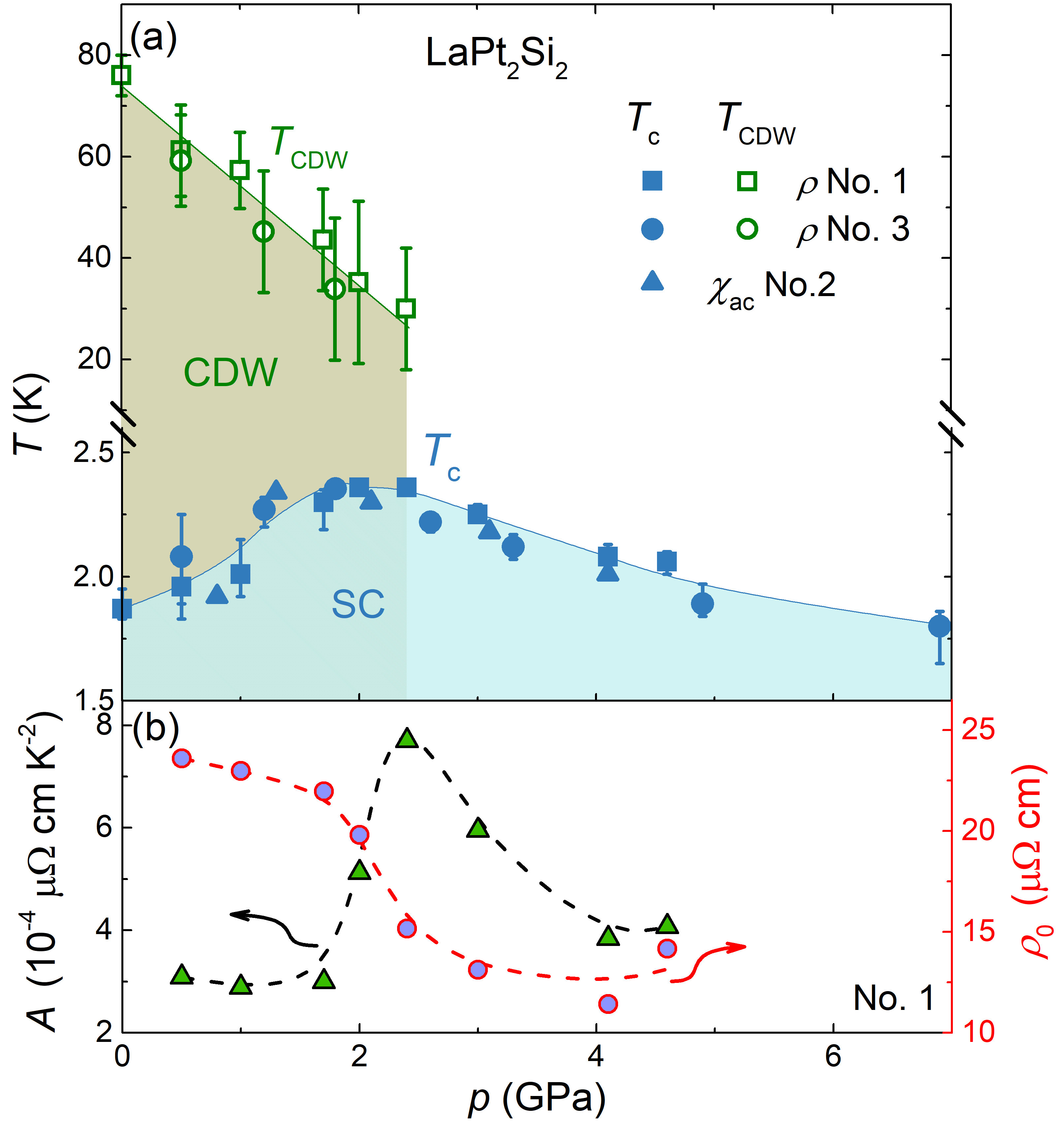}
	\vspace{-12pt} \caption{\label{Figure5}(color online)   (a) Temperature-pressure phase diagram of LaPt$_2$Si$_2$.  $T_{\rm{CDW}}$ is determined from the derivative of the resistivity of two samples, where the error bars in $T_{\rm{CDW}}$ indicate the full width at half minimum of d$\rho$/d$T$. $T_{\rm{c}}$ from $\rho(T)$ corresponds to the temperature where there is a drop to 50~\% of the normal state value. For the  ac susceptibility,  $T_{\rm{c}}$  is the temperature of the onset of the transition. The error bars in the  $T_{\rm{c}}$ from the  resistivity represents where there is a  10~\% and 90~\% drop of the resistivity. (b) Pressure dependence of the $A$ coefficient of the $T^2$ term in $\rho(T)$ and residual resistivity $\rho_0$ of sample No. 1.}
	\vspace{-12pt}
\end{figure}
      
From the electrical transport and magnetic susceptibility measurements under pressure, we constructed  the temperature-pressure phase diagram of LaPt$_2$Si$_2$, which is displayed in Fig. \ref{Figure5}(a). As illustrated in the phase diagram,  the CDW order shifts to lower temperature with the application of hydrostatic pressure, before  suddenly disappearing above around 2.4~GPa. On the other hand, $T_{\rm{c}}$ initially  increases with pressure, reaching a maximum value at around the pressure where CDW order vanishes. Moreover, the dome-like shape of the superconducting state is different from many other examples of SC competing with CDW order, where $T_{\rm{c}}$ often suddenly jumps upon suppressing the CDW transition \cite{Gabovich_2001, Sulfur,  ZrTa3}. Figure \ref{Figure5}(b) shows the results of fitting the resistivity of sample No. 1 where it can be observed that the $A$ coefficient reaches a maximum at around 2.4~GPa, close to the  pressure at which the CDW transition abruptly disappears. Furthermore, $\rho_0$ abruptly decreases upon the suppression  of  CDW order, which may be related to the closure of the CDW gap.

\section{\uppercase\expandafter{\romannumeral4}. DISCUSSION}

  The sudden disappearance of CDW order above 2.4~GPa in LaPt$_2$Si$_2$ indicates that there is a first-order transition, suggesting the lack of a QCP, which can account for the absence of non-Fermi liquid behavior across the phase diagram. 

  As displayed in Fig. \ref{Figure5}, both the  $A$-coefficient and $T_{\rm{c}}$ have maximum values at around the pressure where CDW order disappears. Since, $A$ $\propto$ $N^2(E_F)$ \cite{A_FL}, this suggests that there is a maximum value of $N(E_F)$ at this pressure. Together with the evidence  that LaPt$_2$Si$_2$ is a conventional electron-phonon mediated superconductor \cite{2018uSR,Gupta_2017},  then a  peak in $N(E_F)$ under pressure can lead to a superconducting dome,  since  $T_{\rm{c}}$ $\propto$ $\omega_D$exp[-1/$N(E_F)V$)] in  BCS theory, where  $\omega_D$ is a phonon frequency, and $V$ is the electron pairing potential.   Note that  $\omega_D$ and $V$ generally exhibit a more moderate pressure dependence compared to $N(E_F)$ \cite{1992EPL}. Therefore, our results suggest that the evolution of $T_{\rm{c}}$ under pressure in LaPt$_2$Si$_2$ is likely driven by the variation of $N(E_F)$.  However, further studies are desirable in order to elucidate the nature of the superconducting state and its pressure evolution.

 Below 2.4~GPa , the increase of $N(E_F)$ with increasing pressure can be naturally explained by the suppression of the CDW gap. Above 2.4~GPa, a possible reason for the decrease of  $N(E_F)$ is due to  band broadening upon compressing the lattice \cite{hamlinSC,monteverde2001pressure}. To confirm this,  electronic structure calculations at high pressures are necessary.

It is noted that LaPt$_2$Ge$_2$ is also a CDW superconductor, with a lower $T_{\rm{c}}$ = 0.4~K and a higher $T_{\rm{CDW}}$ = 385~K compared to LaPt$_2$Si$_2$. The latter is accompanied by a structural phase transition from  the tetragonal CaBe$_2$Ge$_2$-type to a monoclinic structure \cite{LPG_SC,PRBLaPt2Ge2}. Moreover, the Fermi surface of LaPt$_2$Ge$_2$ in the tetragonal phase resembles that of LaPt$_2$Si$_2$ \cite{2015FSSciRep,PRBLaPt2Ge2}. Upon varying the ratio of Pt and Ge, it was suggested from NMR measurements that  enhanced structural fluctuations in LaPt$_{2-x}$Ge$_{2+x}$ can possibly give rise to the increase of $T_{\rm{c}}$ \cite{PRBLaPt2Ge2}.  As such, it would also be of interest to look for the role played by structural fluctuations in influencing the superconductivity of LaPt$_2$Si$_2$, which may be addressed by NMR measurements under pressure.

\section{\uppercase\expandafter{\romannumeral5}. CONCLUSION}

In conclusion, we have determined the pressure-temperature phase diagram of LaPt$_2$Si$_2$, which exhibits both superconductivity and CDW order. With the application of hydrostatic pressure, the CDW order is suppressed to lower temperatures  before abruptly vanishing above 2.4~GPa, while $T_{\rm{c}}$ shows a dome-like shape with a maximum value at around the same pressure. Furthermore, we suggest that  the change of $N(E_F)$ under pressure might account for the SC dome in  LaPt$_2$Si$_2$ in the framework of BCS theory. Finally, experiments under pressure, such as NMR, and calculations of the electronic structure of LaPt$_2$Si$_2$ would be useful to gain further understanding of the interplay between CDW order and superconductivity in this system.

\section{ACKNOWLEDGMENTS}

ZH thanks Ritu Gupta for discussion and initial characterization of the sample. This work was supported by the National Key R\&D Program of China (Grants No. 2017YFA0303100 and No. 2016YFA0300202), the National Natural Science Foundation of China (Grants No. U1632275 and No. 11974306), and the Science Challenge Project of China (Grant No. TZ2016004).


\begin{thebibliography}{35}%
	\makeatletter
	\providecommand \@ifxundefined [1]{%
		\@ifx{#1\undefined}
	}%
	\providecommand \@ifnum [1]{%
		\ifnum #1\expandafter \@firstoftwo
		\else \expandafter \@secondoftwo
		\fi
	}%
	\providecommand \@ifx [1]{%
		\ifx #1\expandafter \@firstoftwo
		\else \expandafter \@secondoftwo
		\fi
	}%
	\providecommand \natexlab [1]{#1}%
	\providecommand \enquote  [1]{``#1''}%
	\providecommand \bibnamefont  [1]{#1}%
	\providecommand \bibfnamefont [1]{#1}%
	\providecommand \citenamefont [1]{#1}%
	\providecommand \href@noop [0]{\@secondoftwo}%
	\providecommand \href [0]{\begingroup \@sanitize@url \@href}%
	\providecommand \@href[1]{\@@startlink{#1}\@@href}%
	\providecommand \@@href[1]{\endgroup#1\@@endlink}%
	\providecommand \@sanitize@url [0]{\catcode `\\12\catcode `\$12\catcode
		`\&12\catcode `\#12\catcode `\^12\catcode `\_12\catcode `\%12\relax}%
	\providecommand \@@startlink[1]{}%
	\providecommand \@@endlink[0]{}%
	\providecommand \url  [0]{\begingroup\@sanitize@url \@url }%
	\providecommand \@url [1]{\endgroup\@href {#1}{\urlprefix }}%
	\providecommand \urlprefix  [0]{URL }%
	\providecommand \Eprint [0]{\href }%
	\providecommand \doibase [0]{http://dx.doi.org/}%
	\providecommand \selectlanguage [0]{\@gobble}%
	\providecommand \bibinfo  [0]{\@secondoftwo}%
	\providecommand \bibfield  [0]{\@secondoftwo}%
	\providecommand \translation [1]{[#1]}%
	\providecommand \BibitemOpen [0]{}%
	\providecommand \bibitemStop [0]{}%
	\providecommand \bibitemNoStop [0]{.\EOS\space}%
	\providecommand \EOS [0]{\spacefactor3000\relax}%
	\providecommand \BibitemShut  [1]{\csname bibitem#1\endcsname}%
	\let\auto@bib@innerbib\@empty
	\bibitem [{\citenamefont {Morosan}\ \emph {et~al.}(2006)\citenamefont
		{Morosan}, \citenamefont {Zandbergen}, \citenamefont {Dennis}, \citenamefont
		{Bos}, \citenamefont {Onose}, \citenamefont {Klimczuk}, \citenamefont
		{Ramirez}, \citenamefont {Ong},\ and\ \citenamefont {Cava}}]{NP_CuTS}%
	\BibitemOpen
	\bibfield  {author} {\bibinfo {author} {\bibfnamefont {E.}~\bibnamefont
			{Morosan}}, \bibinfo {author} {\bibfnamefont {H.~W.}\ \bibnamefont
			{Zandbergen}}, \bibinfo {author} {\bibfnamefont {B.~S.}\ \bibnamefont
			{Dennis}}, \bibinfo {author} {\bibfnamefont {J.~W.~G.}\ \bibnamefont {Bos}},
		\bibinfo {author} {\bibfnamefont {Y.}~\bibnamefont {Onose}}, \bibinfo
		{author} {\bibfnamefont {T.}~\bibnamefont {Klimczuk}}, \bibinfo {author}
		{\bibfnamefont {A.~P.}\ \bibnamefont {Ramirez}}, \bibinfo {author}
		{\bibfnamefont {N.~P.}\ \bibnamefont {Ong}}, \ and\ \bibinfo {author}
		{\bibfnamefont {R.~J.}\ \bibnamefont {Cava}},\ }\bibfield  {title} {\enquote
		{\bibinfo {title} {Superconductivity in $\mathrm{Cu_xTiSe_2}$},}\ }\href
	{\doibase 10.1038/nphys360} {\bibfield  {journal} {\bibinfo  {journal}
			{Nature Phys.}\ }\textbf {\bibinfo {volume} {2}},\ \bibinfo {pages}
		{544--550} (\bibinfo {year} {2006})}\BibitemShut {NoStop}%
	\bibitem [{\citenamefont {Kusmartseva}\ \emph {et~al.}(2009)\citenamefont
		{Kusmartseva}, \citenamefont {Sipos}, \citenamefont {Berger}, \citenamefont
		{Forr\'o},\ and\ \citenamefont {Tuti\ifmmode~\check{s}\else
			\v{s}\fi{}}}]{Pressure_TS}%
	\BibitemOpen
	\bibfield  {author} {\bibinfo {author} {\bibfnamefont {A.~F.}\ \bibnamefont
			{Kusmartseva}}, \bibinfo {author} {\bibfnamefont {B.}~\bibnamefont {Sipos}},
		\bibinfo {author} {\bibfnamefont {H.}~\bibnamefont {Berger}}, \bibinfo
		{author} {\bibfnamefont {L.}~\bibnamefont {Forr\'o}}, \ and\ \bibinfo
		{author} {\bibfnamefont {E.}~\bibnamefont {Tuti\ifmmode~\check{s}\else
				\v{s}\fi{}}},\ }\bibfield  {title} {\enquote {\bibinfo {title} {Pressure
				induced superconductivity in pristine 1${T}$-$\mathrm{TiSe_2}$},}\ }\href
	{\doibase 10.1103/PhysRevLett.103.236401} {\bibfield  {journal} {\bibinfo
			{journal} {Phys. Rev. Lett.}\ }\textbf {\bibinfo {volume} {103}},\ \bibinfo
		{pages} {236401} (\bibinfo {year} {2009})}\BibitemShut {NoStop}%
	\bibitem [{\citenamefont {Gruner}\ \emph {et~al.}(2017)\citenamefont {Gruner},
		\citenamefont {Jang}, \citenamefont {Huesges}, \citenamefont {Cardoso-Gil},
		\citenamefont {Fecher}, \citenamefont {Koza}, \citenamefont {Stockert},
		\citenamefont {Mackenzie}, \citenamefont {Brando}, ,\ and\ \citenamefont
		{Geibel}}]{NP2017_LPI}%
	\BibitemOpen
	\bibfield  {author} {\bibinfo {author} {\bibfnamefont {T.}~\bibnamefont
			{Gruner}}, \bibinfo {author} {\bibfnamefont {D.~J.}\ \bibnamefont {Jang}},
		\bibinfo {author} {\bibfnamefont {Z.}~\bibnamefont {Huesges}}, \bibinfo
		{author} {\bibfnamefont {R.}~\bibnamefont {Cardoso-Gil}}, \bibinfo {author}
		{\bibfnamefont {G.~H.}\ \bibnamefont {Fecher}}, \bibinfo {author}
		{\bibfnamefont {M.~M.}\ \bibnamefont {Koza}}, \bibinfo {author}
		{\bibfnamefont {O.}~\bibnamefont {Stockert}}, \bibinfo {author}
		{\bibfnamefont {A.~P.}\ \bibnamefont {Mackenzie}}, \bibinfo {author}
		{\bibfnamefont {M.}~\bibnamefont {Brando}}, , \ and\ \bibinfo {author}
		{\bibfnamefont {C.}~\bibnamefont {Geibel}},\ }\bibfield  {title} {\enquote
		{\bibinfo {title} {Charge density wave quantum critical point with strong
				enhancement of superconductivity},}\ }\href {\doibase 10.1038/nphys4191}
	{\bibfield  {journal} {\bibinfo  {journal} {Nature Phys.}\ }\textbf {\bibinfo
			{volume} {13}},\ \bibinfo {pages} {967--972} (\bibinfo {year}
		{2017})}\BibitemShut {NoStop}%
	\bibitem [{\citenamefont {Testardi}(1975)}]{RMPA15}%
	\BibitemOpen
	\bibfield  {author} {\bibinfo {author} {\bibfnamefont {L.~R.}\ \bibnamefont
			{Testardi}},\ }\bibfield  {title} {\enquote {\bibinfo {title} {Structural
				instability and superconductivity in ${A}$-15 compounds},}\ }\href {\doibase
		10.1103/RevModPhys.47.637} {\bibfield  {journal} {\bibinfo  {journal} {Rev.
				Mod. Phys.}\ }\textbf {\bibinfo {volume} {47}},\ \bibinfo {pages} {637--648}
		(\bibinfo {year} {1975})}\BibitemShut {NoStop}%
	\bibitem [{\citenamefont {Coleman}\ \emph {et~al.}(1973)\citenamefont
		{Coleman}, \citenamefont {Cohen}, \citenamefont {Sandman}, \citenamefont
		{Yamagishi}, \citenamefont {Garito},\ and\ \citenamefont
		{Heeger}}]{COLEMAN1973}%
	\BibitemOpen
	\bibfield  {author} {\bibinfo {author} {\bibfnamefont {L.~B.}\ \bibnamefont
			{Coleman}}, \bibinfo {author} {\bibfnamefont {M.~J.}\ \bibnamefont {Cohen}},
		\bibinfo {author} {\bibfnamefont {D.~J.}\ \bibnamefont {Sandman}}, \bibinfo
		{author} {\bibfnamefont {F.~G.}\ \bibnamefont {Yamagishi}}, \bibinfo {author}
		{\bibfnamefont {A.~F.}\ \bibnamefont {Garito}}, \ and\ \bibinfo {author}
		{\bibfnamefont {A.~J.}\ \bibnamefont {Heeger}},\ }\bibfield  {title}
	{\enquote {\bibinfo {title} {Superconducting fluctuations and the peierls
				instability in an organic solid},}\ }\href {\doibase
		https://doi.org/10.1016/0038-1098(73)90127-0} {\bibfield  {journal} {\bibinfo
			{journal} {Solid State Commun.}\ }\textbf {\bibinfo {volume} {12}},\
		\bibinfo {pages} {1125 -- 1132} (\bibinfo {year} {1973})}\BibitemShut
	{NoStop}%
	\bibitem [{\citenamefont {Wilson}\ \emph {et~al.}(1975)\citenamefont {Wilson},
		\citenamefont {Di~Salvo},\ and\ \citenamefont {Mahajan}}]{CDW1975}%
	\BibitemOpen
	\bibfield  {author} {\bibinfo {author} {\bibfnamefont {J.~A.}\ \bibnamefont
			{Wilson}}, \bibinfo {author} {\bibfnamefont {F.~J.}\ \bibnamefont
			{Di~Salvo}}, \ and\ \bibinfo {author} {\bibfnamefont {S.}~\bibnamefont
			{Mahajan}},\ }\bibfield  {title} {\enquote {\bibinfo {title} {Charge-density
				waves and superlattices in the metallic layered transition metal
				dichalcogenides},}\ }\href {\doibase 10.1080/00018737500101391} {\bibfield
		{journal} {\bibinfo  {journal} {Adv. Phys.}\ }\textbf {\bibinfo {volume}
			{24}},\ \bibinfo {pages} {117--201} (\bibinfo {year} {1975})}\BibitemShut
	{NoStop}%
	\bibitem [{\citenamefont {Edwards}\ \emph {et~al.}(1994)\citenamefont
		{Edwards}, \citenamefont {Barr}, \citenamefont {Markert},\ and\ \citenamefont
		{de~Lozanne}}]{CDWCu}%
	\BibitemOpen
	\bibfield  {author} {\bibinfo {author} {\bibfnamefont {H.~L.}\ \bibnamefont
			{Edwards}}, \bibinfo {author} {\bibfnamefont {A.~L.}\ \bibnamefont {Barr}},
		\bibinfo {author} {\bibfnamefont {J.~T.}\ \bibnamefont {Markert}}, \ and\
		\bibinfo {author} {\bibfnamefont {A.~L.}\ \bibnamefont {de~Lozanne}},\
	}\bibfield  {title} {\enquote {\bibinfo {title} {Modulations in the $\mathrm
				{CuO}$ chain layer of $\mathrm{YBa_2Cu_3O_{7-\delta}}$: Charge density
				waves?}}\ }\href {\doibase 10.1103/PhysRevLett.73.1154} {\bibfield  {journal}
		{\bibinfo  {journal} {Phys. Rev. Lett.}\ }\textbf {\bibinfo {volume} {73}},\
		\bibinfo {pages} {1154--1157} (\bibinfo {year} {1994})}\BibitemShut {NoStop}%
	\bibitem [{\citenamefont {Gabovich}\ \emph {et~al.}(2002)\citenamefont
		{Gabovich}, \citenamefont {Voitenko},\ and\ \citenamefont
		{Ausloos}}]{Review2002}%
	\BibitemOpen
	\bibfield  {author} {\bibinfo {author} {\bibfnamefont {A.~M.}\ \bibnamefont
			{Gabovich}}, \bibinfo {author} {\bibfnamefont {A.~M.}\ \bibnamefont
			{Voitenko}}, \ and\ \bibinfo {author} {\bibfnamefont {M.}~\bibnamefont
			{Ausloos}},\ }\bibfield  {title} {\enquote {\bibinfo {title} {Charge- and
				spin-density waves in existing superconductors: competition between cooper
				pairing and peierls or excitonic instabilities},}\ }\href {\doibase
		https://doi.org/10.1016/S0370-1573(02)00029-7} {\bibfield  {journal}
		{\bibinfo  {journal} {Physics Reports}\ }\textbf {\bibinfo {volume} {367}},\
		\bibinfo {pages} {583 -- 709} (\bibinfo {year} {2002})}\BibitemShut {NoStop}%
	\bibitem [{\citenamefont {Gabovich}\ \emph {et~al.}(2001)\citenamefont
		{Gabovich}, \citenamefont {Voitenko}, \citenamefont {Annett},\ and\
		\citenamefont {Ausloos}}]{Gabovich_2001}%
	\BibitemOpen
	\bibfield  {author} {\bibinfo {author} {\bibfnamefont {A.~M.}\ \bibnamefont
			{Gabovich}}, \bibinfo {author} {\bibfnamefont {A.~I.}\ \bibnamefont
			{Voitenko}}, \bibinfo {author} {\bibfnamefont {J.~F.}\ \bibnamefont
			{Annett}}, \ and\ \bibinfo {author} {\bibfnamefont {M.}~\bibnamefont
			{Ausloos}},\ }\bibfield  {title} {\enquote {\bibinfo {title} {Charge- and
				spin-density-wave superconductors},}\ }\href {\doibase
		10.1088/0953-2048/14/4/201} {\bibfield  {journal} {\bibinfo  {journal}
			{Superconductor Science and Technology}\ }\textbf {\bibinfo {volume} {14}},\
		\bibinfo {pages} {R1--R27} (\bibinfo {year} {2001})}\BibitemShut {NoStop}%
	\bibitem [{\citenamefont {Freitas}\ \emph {et~al.}(2016)\citenamefont
		{Freitas}, \citenamefont {Rodi\`ere}, \citenamefont {Osorio}, \citenamefont
		{Navarro-Moratalla}, \citenamefont {Nemes}, \citenamefont {Tissen},
		\citenamefont {Cario}, \citenamefont {Coronado}, \citenamefont
		{Garc\'{\i}a-Hern\'andez}, \citenamefont {Vieira}, \citenamefont {N\'u\~nez
			Regueiro},\ and\ \citenamefont {Suderow}}]{Pressure_TaS}%
	\BibitemOpen
	\bibfield  {author} {\bibinfo {author} {\bibfnamefont {D.~C.}\ \bibnamefont
			{Freitas}}, \bibinfo {author} {\bibfnamefont {P.}~\bibnamefont {Rodi\`ere}},
		\bibinfo {author} {\bibfnamefont {M.~R.}\ \bibnamefont {Osorio}}, \bibinfo
		{author} {\bibfnamefont {E.}~\bibnamefont {Navarro-Moratalla}}, \bibinfo
		{author} {\bibfnamefont {N.~M.}\ \bibnamefont {Nemes}}, \bibinfo {author}
		{\bibfnamefont {V.~G.}\ \bibnamefont {Tissen}}, \bibinfo {author}
		{\bibfnamefont {L.}~\bibnamefont {Cario}}, \bibinfo {author} {\bibfnamefont
			{E.}~\bibnamefont {Coronado}}, \bibinfo {author} {\bibfnamefont
			{M.}~\bibnamefont {Garc\'{\i}a-Hern\'andez}}, \bibinfo {author}
		{\bibfnamefont {S.}~\bibnamefont {Vieira}}, \bibinfo {author} {\bibfnamefont
			{M.}~\bibnamefont {N\'u\~nez-Regueiro}}, \ and\ \bibinfo {author}
		{\bibfnamefont {H.}~\bibnamefont {Suderow}},\ }\bibfield  {title} {\enquote
		{\bibinfo {title} {Strong enhancement of superconductivity at high pressures
				within the charge-density-wave states of 2${H}$-$\mathrm{TaS_2}$ and
				2${H}$-$\mathrm{TaSe_2}$},}\ }\href {\doibase 10.1103/PhysRevB.93.184512}
	{\bibfield  {journal} {\bibinfo  {journal} {Phys. Rev. B}\ }\textbf {\bibinfo
			{volume} {93}},\ \bibinfo {pages} {184512} (\bibinfo {year}
		{2016})}\BibitemShut {NoStop}%
	\bibitem [{\citenamefont {Bhoi}\ \emph {et~al.}(2016)\citenamefont {Bhoi},
		\citenamefont {Khim}, \citenamefont {Nam}, \citenamefont {Lee}, \citenamefont
		{Kim}, \citenamefont {Jeon}, \citenamefont {Min}, \citenamefont {Park},\ and\
		\citenamefont {Kim}}]{Pressure_PdTaSe}%
	\BibitemOpen
	\bibfield  {author} {\bibinfo {author} {\bibfnamefont {D.}~\bibnamefont
			{Bhoi}}, \bibinfo {author} {\bibfnamefont {S.}~\bibnamefont {Khim}}, \bibinfo
		{author} {\bibfnamefont {W.}~\bibnamefont {Nam}}, \bibinfo {author}
		{\bibfnamefont {B.~S.}\ \bibnamefont {Lee}}, \bibinfo {author} {\bibfnamefont
			{C.}~\bibnamefont {Kim}}, \bibinfo {author} {\bibfnamefont {B.~G.}\
			\bibnamefont {Jeon}}, \bibinfo {author} {\bibfnamefont {B.~H.}\ \bibnamefont
			{Min}}, \bibinfo {author} {\bibfnamefont {S.}~\bibnamefont {Park}}, \ and\
		\bibinfo {author} {\bibfnamefont {K.~H.}\ \bibnamefont {Kim}},\ }\bibfield
	{title} {\enquote {\bibinfo {title} {Interplay of charge density wave and
				multiband superconductivity in 2${H}$-$\mathrm{Pd_xTaSe_2}$},}\ }\href
	{\doibase 10.1038/srep24068} {\bibfield  {journal} {\bibinfo  {journal}
			{Scientific Reports}\ }\textbf {\bibinfo {volume} {6}},\ \bibinfo {pages}
		{24068} (\bibinfo {year} {2016})}\BibitemShut {NoStop}%
	\bibitem [{\citenamefont {Monteverde}\ \emph {et~al.}(2013)\citenamefont
		{Monteverde}, \citenamefont {Lorenzana}, \citenamefont {Monceau},\ and\
		\citenamefont {N\'u\~nez{-}Regueiro}}]{Pressure_TaS3}%
	\BibitemOpen
	\bibfield  {author} {\bibinfo {author} {\bibfnamefont {M.}~\bibnamefont
			{Monteverde}}, \bibinfo {author} {\bibfnamefont {J.}~\bibnamefont
			{Lorenzana}}, \bibinfo {author} {\bibfnamefont {P.}~\bibnamefont {Monceau}},
		\ and\ \bibinfo {author} {\bibfnamefont {M.}~\bibnamefont
			{N\'u\~nez{-}Regueiro}},\ }\bibfield  {title} {\enquote {\bibinfo {title}
			{Quantum critical point and superconducting dome in the pressure phase
				diagram of $o$-$\mathrm{TaS_3}$},}\ }\href {\doibase
		10.1103/PhysRevB.88.180504} {\bibfield  {journal} {\bibinfo  {journal} {Phys.
				Rev. B}\ }\textbf {\bibinfo {volume} {88}},\ \bibinfo {pages} {180504(R)}
		(\bibinfo {year} {2013})}\BibitemShut {NoStop}%
	\bibitem [{\citenamefont {Klintberg}\ \emph {et~al.}(2012)\citenamefont
		{Klintberg}, \citenamefont {Goh}, \citenamefont {Alireza}, \citenamefont
		{Saines}, \citenamefont {Tompsett}, \citenamefont {Logg}, \citenamefont
		{Yang}, \citenamefont {Chen}, \citenamefont {Yoshimura},\ and\ \citenamefont
		{Grosche}}]{Malte2012CIS}%
	\BibitemOpen
	\bibfield  {author} {\bibinfo {author} {\bibfnamefont {L.~E.}\ \bibnamefont
			{Klintberg}}, \bibinfo {author} {\bibfnamefont {S.~K.}\ \bibnamefont {Goh}},
		\bibinfo {author} {\bibfnamefont {P.~L.}\ \bibnamefont {Alireza}}, \bibinfo
		{author} {\bibfnamefont {P.~J.}\ \bibnamefont {Saines}}, \bibinfo {author}
		{\bibfnamefont {D.~A.}\ \bibnamefont {Tompsett}}, \bibinfo {author}
		{\bibfnamefont {P.~W.}\ \bibnamefont {Logg}}, \bibinfo {author}
		{\bibfnamefont {J.~H.}\ \bibnamefont {Yang}}, \bibinfo {author}
		{\bibfnamefont {B.}~\bibnamefont {Chen}}, \bibinfo {author} {\bibfnamefont
			{K.}~\bibnamefont {Yoshimura}}, \ and\ \bibinfo {author} {\bibfnamefont
			{F.~M.}\ \bibnamefont {Grosche}},\ }\bibfield  {title} {\enquote {\bibinfo
			{title} {Pressure- and composition-induced structural quantum phase
				transition in the cubic superconductor $\mathrm{(Ca,Sr)_3Ir_4Sn_{13}}$},}\
	}\href {\doibase 10.1103/PhysRevLett.109.237008} {\bibfield  {journal}
		{\bibinfo  {journal} {Phys. Rev. Lett.}\ }\textbf {\bibinfo {volume} {109}},\
		\bibinfo {pages} {237008} (\bibinfo {year} {2012})}\BibitemShut {NoStop}%
	\bibitem [{\citenamefont {Stewart}(2017)}]{UncoSC_Stewart}%
	\BibitemOpen
	\bibfield  {author} {\bibinfo {author} {\bibfnamefont {G.~R.}\ \bibnamefont
			{Stewart}},\ }\bibfield  {title} {\enquote {\bibinfo {title} {Unconventional
				superconductivity},}\ }\href {\doibase 10.1080/00018732.2017.1331615}
	{\bibfield  {journal} {\bibinfo  {journal} {Advances in Physics}\ }\textbf
		{\bibinfo {volume} {66}},\ \bibinfo {pages} {75--196} (\bibinfo {year}
		{2017})}\BibitemShut {NoStop}%
	\bibitem [{\citenamefont {Grube}\ \emph {et~al.}(2017)\citenamefont {Grube},
		\citenamefont {Zaum}, \citenamefont {Stockert},\ and\ \citenamefont
		{Si}}]{NP2017}%
	\BibitemOpen
	\bibfield  {author} {\bibinfo {author} {\bibfnamefont {K.}~\bibnamefont
			{Grube}}, \bibinfo {author} {\bibfnamefont {S.}~\bibnamefont {Zaum}},
		\bibinfo {author} {\bibfnamefont {O.}~\bibnamefont {Stockert}}, \ and\
		\bibinfo {author} {\bibfnamefont {Q.}~\bibnamefont {Si}},\ }\bibfield
	{title} {\enquote {\bibinfo {title} {Multidimensional entropy landscape of
				quantum criticality},}\ }\href {\doibase 10.1038/nphys4113} {\bibfield
		{journal} {\bibinfo  {journal} {Nature Phys.}\ }\textbf {\bibinfo {volume}
			{13}},\ \bibinfo {pages} {742--745} (\bibinfo {year} {2017})}\BibitemShut
	{NoStop}%
	\bibitem [{\citenamefont {Joe}\ \emph {et~al.}(2014)\citenamefont {Joe},
		\citenamefont {Chen}, \citenamefont {Ghaemi}, \citenamefont {Finkelstein},
		\citenamefont {de~la Peña}, \citenamefont {Gan}, \citenamefont {Lee},
		\citenamefont {Yuan}, \citenamefont {Geck}, \citenamefont {MacDougall},
		\citenamefont {Chiang}, \citenamefont {Cooper}, \citenamefont {Fradkin},\
		and\ \citenamefont {Abbamonte}}]{NP_TS_Domain}%
	\BibitemOpen
	\bibfield  {author} {\bibinfo {author} {\bibfnamefont {Y.~I.}\ \bibnamefont
			{Joe}}, \bibinfo {author} {\bibfnamefont {X.~M.}\ \bibnamefont {Chen}},
		\bibinfo {author} {\bibfnamefont {P.}~\bibnamefont {Ghaemi}}, \bibinfo
		{author} {\bibfnamefont {K.~D.}\ \bibnamefont {Finkelstein}}, \bibinfo
		{author} {\bibfnamefont {G.~A.}\ \bibnamefont {de~la Peña}}, \bibinfo
		{author} {\bibfnamefont {Y.}~\bibnamefont {Gan}}, \bibinfo {author}
		{\bibfnamefont {J.~C.~T.}\ \bibnamefont {Lee}}, \bibinfo {author}
		{\bibfnamefont {S.}~\bibnamefont {Yuan}}, \bibinfo {author} {\bibfnamefont
			{J.}~\bibnamefont {Geck}}, \bibinfo {author} {\bibfnamefont {G.~J.}\
			\bibnamefont {MacDougall}}, \bibinfo {author} {\bibfnamefont {T.~C.}\
			\bibnamefont {Chiang}}, \bibinfo {author} {\bibfnamefont {S.~L.}\
			\bibnamefont {Cooper}}, \bibinfo {author} {\bibfnamefont {E.}~\bibnamefont
			{Fradkin}}, \ and\ \bibinfo {author} {\bibfnamefont {P.}~\bibnamefont
			{Abbamonte}},\ }\bibfield  {title} {\enquote {\bibinfo {title} {Emergence of
				charge density wave domain walls above the superconducting dome in
				1${T}$-$\mathrm{TiSe_2}$},}\ }\href {\doibase 10.1038/nphys2935} {\bibfield
		{journal} {\bibinfo  {journal} {Nature Phys.}\ }\textbf {\bibinfo {volume}
			{10}},\ \bibinfo {pages} {421--425} (\bibinfo {year} {2014})}\BibitemShut
	{NoStop}%
	\bibitem [{\citenamefont {Fang}\ \emph {et~al.}(2012)\citenamefont {Fang},
		\citenamefont {Dong}, \citenamefont {Wang}, \citenamefont {Chen},
		\citenamefont {Cheng}, \citenamefont {Shi}, \citenamefont {Zheng},
		\citenamefont {Xu}, \citenamefont {Wang}, \citenamefont {Li},\ and\
		\citenamefont {Wang}}]{Sr122}%
	\BibitemOpen
	\bibfield  {author} {\bibinfo {author} {\bibfnamefont {A.~F.}\ \bibnamefont
			{Fang}}, \bibinfo {author} {\bibfnamefont {T.}~\bibnamefont {Dong}}, \bibinfo
		{author} {\bibfnamefont {H.~P.}\ \bibnamefont {Wang}}, \bibinfo {author}
		{\bibfnamefont {Z.~G.}\ \bibnamefont {Chen}}, \bibinfo {author}
		{\bibfnamefont {B.}~\bibnamefont {Cheng}}, \bibinfo {author} {\bibfnamefont
			{Y.~G.}\ \bibnamefont {Shi}}, \bibinfo {author} {\bibfnamefont
			{P.}~\bibnamefont {Zheng}}, \bibinfo {author} {\bibfnamefont
			{G.}~\bibnamefont {Xu}}, \bibinfo {author} {\bibfnamefont {L.}~\bibnamefont
			{Wang}}, \bibinfo {author} {\bibfnamefont {J.~Q.}\ \bibnamefont {Li}}, \ and\
		\bibinfo {author} {\bibfnamefont {N.~L.}\ \bibnamefont {Wang}},\ }\bibfield
	{title} {\enquote {\bibinfo {title} {Single-crystal growth and optical
				conductivity of $\mathrm{SrPt_2As_2}$ superconductors},}\ }\href {\doibase
		10.1103/PhysRevB.85.184520} {\bibfield  {journal} {\bibinfo  {journal} {Phys.
				Rev. B}\ }\textbf {\bibinfo {volume} {85}},\ \bibinfo {pages} {184520}
		(\bibinfo {year} {2012})}\BibitemShut {NoStop}%
	\bibitem [{\citenamefont {Jiang}\ \emph {et~al.}(2014)\citenamefont {Jiang},
		\citenamefont {Guo}, \citenamefont {Weng}, \citenamefont {Wang},
		\citenamefont {Chen}, \citenamefont {Chen}, \citenamefont {Pang},
		\citenamefont {Shang}, \citenamefont {Lu},\ and\ \citenamefont
		{Yuan}}]{jiang2014}%
	\BibitemOpen
	\bibfield  {author} {\bibinfo {author} {\bibfnamefont {W.~B.}\ \bibnamefont
			{Jiang}}, \bibinfo {author} {\bibfnamefont {C.~Y.}\ \bibnamefont {Guo}},
		\bibinfo {author} {\bibfnamefont {Z.~F.}\ \bibnamefont {Weng}}, \bibinfo
		{author} {\bibfnamefont {Y.~F.}\ \bibnamefont {Wang}}, \bibinfo {author}
		{\bibfnamefont {Y.~H.}\ \bibnamefont {Chen}}, \bibinfo {author}
		{\bibfnamefont {Y.}~\bibnamefont {Chen}}, \bibinfo {author} {\bibfnamefont
			{G.~M.}\ \bibnamefont {Pang}}, \bibinfo {author} {\bibfnamefont
			{T.}~\bibnamefont {Shang}}, \bibinfo {author} {\bibfnamefont
			{T.}~\bibnamefont {Lu}}, \ and\ \bibinfo {author} {\bibfnamefont {H.~Q.}\
			\bibnamefont {Yuan}},\ }\bibfield  {title} {\enquote {\bibinfo {title}
			{Superconductivity and structural distortion in $\mathrm{BaPt_2As_2}$},}\
	}\href {\doibase 10.1088/0953-8984/27/2/022202} {\bibfield  {journal}
		{\bibinfo  {journal} {J. Phys.: Condens. Matter}\ }\textbf {\bibinfo {volume}
			{27}},\ \bibinfo {pages} {022202} (\bibinfo {year} {2014})}\BibitemShut
	{NoStop}%
	\bibitem [{\citenamefont {Guo}\ \emph {et~al.}(2016)\citenamefont {Guo},
		\citenamefont {Jiang}, \citenamefont {Smidman}, \citenamefont {Han},
		\citenamefont {Malliakas}, \citenamefont {Shen}, \citenamefont {Wang},
		\citenamefont {Chen}, \citenamefont {Lu}, \citenamefont {Kanatzidis},\ and\
		\citenamefont {Yuan}}]{GuoBa}%
	\BibitemOpen
	\bibfield  {author} {\bibinfo {author} {\bibfnamefont {C.~Y.}\ \bibnamefont
			{Guo}}, \bibinfo {author} {\bibfnamefont {W.~B.}\ \bibnamefont {Jiang}},
		\bibinfo {author} {\bibfnamefont {M.}~\bibnamefont {Smidman}}, \bibinfo
		{author} {\bibfnamefont {F.}~\bibnamefont {Han}}, \bibinfo {author}
		{\bibfnamefont {C.~D.}\ \bibnamefont {Malliakas}}, \bibinfo {author}
		{\bibfnamefont {B.}~\bibnamefont {Shen}}, \bibinfo {author} {\bibfnamefont
			{Y.~F.}\ \bibnamefont {Wang}}, \bibinfo {author} {\bibfnamefont
			{Y.}~\bibnamefont {Chen}}, \bibinfo {author} {\bibfnamefont {X.}~\bibnamefont
			{Lu}}, \bibinfo {author} {\bibfnamefont {M.~G.}\ \bibnamefont {Kanatzidis}},
		\ and\ \bibinfo {author} {\bibfnamefont {H.~Q.}\ \bibnamefont {Yuan}},\
	}\bibfield  {title} {\enquote {\bibinfo {title} {Superconductivity and
				multiple pressure-induced phases in $\mathrm{BaPt_2As_2}$},}\ }\href
	{\doibase 10.1103/PhysRevB.94.184506} {\bibfield  {journal} {\bibinfo
			{journal} {Phys. Rev. B}\ }\textbf {\bibinfo {volume} {94}},\ \bibinfo
		{pages} {184506} (\bibinfo {year} {2016})}\BibitemShut {NoStop}%
	\bibitem [{\citenamefont {Nagano}\ \emph {et~al.}(2013)\citenamefont {Nagano},
		\citenamefont {Araoka}, \citenamefont {Mitsuda}, \citenamefont {Yayama},
		\citenamefont {Wada}, \citenamefont {Ichihara}, \citenamefont {Isobe},\ and\
		\citenamefont {Ueda}}]{2013CDW}%
	\BibitemOpen
	\bibfield  {author} {\bibinfo {author} {\bibfnamefont {Y.}~\bibnamefont
			{Nagano}}, \bibinfo {author} {\bibfnamefont {N.}~\bibnamefont {Araoka}},
		\bibinfo {author} {\bibfnamefont {A.}~\bibnamefont {Mitsuda}}, \bibinfo
		{author} {\bibfnamefont {H.}~\bibnamefont {Yayama}}, \bibinfo {author}
		{\bibfnamefont {H.}~\bibnamefont {Wada}}, \bibinfo {author} {\bibfnamefont
			{M.}~\bibnamefont {Ichihara}}, \bibinfo {author} {\bibfnamefont {Ma.}\
			\bibnamefont {Isobe}}, \ and\ \bibinfo {author} {\bibfnamefont
			{Y.}~\bibnamefont {Ueda}},\ }\bibfield  {title} {\enquote {\bibinfo {title}
			{Charge density wave and superconductivity of $\mathrm{LaPt_2Si_2}$ ({R} =
				{Y}, {La}, {Nd}, and {Lu})},}\ }\href {\doibase 10.7566/JPSJ.82.064715}
	{\bibfield  {journal} {\bibinfo  {journal} {J. Phys. Soc. Jpn.}\ }\textbf
		{\bibinfo {volume} {82}},\ \bibinfo {pages} {064715} (\bibinfo {year}
		{2013})}\BibitemShut {NoStop}%
	\bibitem [{\citenamefont {Gupta}\ \emph {et~al.}(2017)\citenamefont {Gupta},
		\citenamefont {Dhar}, \citenamefont {Thamizhavel}, \citenamefont {Rajeev},\
		and\ \citenamefont {Hossain}}]{Gupta_2017}%
	\BibitemOpen
	\bibfield  {author} {\bibinfo {author} {\bibfnamefont {R.}~\bibnamefont
			{Gupta}}, \bibinfo {author} {\bibfnamefont {S.~K.}\ \bibnamefont {Dhar}},
		\bibinfo {author} {\bibfnamefont {A.}~\bibnamefont {Thamizhavel}}, \bibinfo
		{author} {\bibfnamefont {K.~P.}\ \bibnamefont {Rajeev}}, \ and\ \bibinfo
		{author} {\bibfnamefont {Z.}~\bibnamefont {Hossain}},\ }\bibfield  {title}
	{\enquote {\bibinfo {title} {Superconducting and charge density wave
				transition in single crystalline $\mathrm{LaPt_2Si_2}$},}\ }\href {\doibase
		10.1088/1361-648x/aa70a7} {\bibfield  {journal} {\bibinfo  {journal} {J.
				Phys.: Condens. Matter}\ }\textbf {\bibinfo {volume} {29}},\ \bibinfo {pages}
		{255601} (\bibinfo {year} {2017})}\BibitemShut {NoStop}%
	\bibitem [{\citenamefont {Gupta}\ \emph {et~al.}(2018)\citenamefont {Gupta},
		\citenamefont {Rajeev},\ and\ \citenamefont {Hossain}}]{2018TT}%
	\BibitemOpen
	\bibfield  {author} {\bibinfo {author} {\bibfnamefont {R.}~\bibnamefont
			{Gupta}}, \bibinfo {author} {\bibfnamefont {K.~P.}\ \bibnamefont {Rajeev}}, \
		and\ \bibinfo {author} {\bibfnamefont {Z.}~\bibnamefont {Hossain}},\
	}\bibfield  {title} {\enquote {\bibinfo {title} {Thermal transport studies on
				charge density wave materials $\mathrm{LaPt_2Si_2}$ and
				$\mathrm{PrPt_2Si_2}$},}\ }\href {\doibase 10.1088/1361-648x/aae766}
	{\bibfield  {journal} {\bibinfo  {journal} {J. Phys.: Condens. Matter}\
		}\textbf {\bibinfo {volume} {30}},\ \bibinfo {pages} {475603} (\bibinfo
		{year} {2018})}\BibitemShut {NoStop}%
	\bibitem [{\citenamefont {Hase}\ and\ \citenamefont
		{Yanagisawa}(2013)}]{2013FS}%
	\BibitemOpen
	\bibfield  {author} {\bibinfo {author} {\bibfnamefont {I.}~\bibnamefont
			{Hase}}\ and\ \bibinfo {author} {\bibfnamefont {T.}~\bibnamefont
			{Yanagisawa}},\ }\bibfield  {title} {\enquote {\bibinfo {title} {Electronic
				structure of $\mathrm{LaPt_2Si_2}$},}\ }\href {\doibase
		https://doi.org/10.1016/j.physc.2012.02.047} {\bibfield  {journal} {\bibinfo
			{journal} {Physica C: Superconductivity}\ }\textbf {\bibinfo {volume}
			{484}},\ \bibinfo {pages} {59 -- 61} (\bibinfo {year} {2013})}\BibitemShut
	{NoStop}%
	\bibitem [{\citenamefont {Kim}\ \emph {et~al.}(2015)\citenamefont {Kim},
		\citenamefont {Kim},\ and\ \citenamefont {Min}}]{2015FSSciRep}%
	\BibitemOpen
	\bibfield  {author} {\bibinfo {author} {\bibfnamefont {S.}~\bibnamefont
			{Kim}}, \bibinfo {author} {\bibfnamefont {K.}~\bibnamefont {Kim}}, \ and\
		\bibinfo {author} {\bibfnamefont {B.~I.}\ \bibnamefont {Min}},\ }\bibfield
	{title} {\enquote {\bibinfo {title} {The mechanism of charge density wave in
				{Pt}-based layered superconductors: $\mathrm{SrPt_2As_2}$ and
				$\mathrm{LaPt_2Si_2}$},}\ }\href {\doibase 10.1038/srep15052} {\bibfield
		{journal} {\bibinfo  {journal} {Sci. Rep.}\ }\textbf {\bibinfo {volume}
			{5}},\ \bibinfo {pages} {15052} (\bibinfo {year} {2015})}\BibitemShut
	{NoStop}%
	\bibitem [{\citenamefont {Aoyama}\ \emph {et~al.}(2018)\citenamefont {Aoyama},
		\citenamefont {Kubo}, \citenamefont {Matsuno}, \citenamefont {Kotegawa},
		\citenamefont {Tou}, \citenamefont {Mitsuda}, \citenamefont {Nagano},
		\citenamefont {Araoka}, \citenamefont {Wada},\ and\ \citenamefont
		{Yamada}}]{2018NMR}%
	\BibitemOpen
	\bibfield  {author} {\bibinfo {author} {\bibfnamefont {T.}~\bibnamefont
			{Aoyama}}, \bibinfo {author} {\bibfnamefont {T.}~\bibnamefont {Kubo}},
		\bibinfo {author} {\bibfnamefont {H.}~\bibnamefont {Matsuno}}, \bibinfo
		{author} {\bibfnamefont {H.}~\bibnamefont {Kotegawa}}, \bibinfo {author}
		{\bibfnamefont {H.}~\bibnamefont {Tou}}, \bibinfo {author} {\bibfnamefont
			{A.}~\bibnamefont {Mitsuda}}, \bibinfo {author} {\bibfnamefont {Yu.}\
			\bibnamefont {Nagano}}, \bibinfo {author} {\bibfnamefont {N.}~\bibnamefont
			{Araoka}}, \bibinfo {author} {\bibfnamefont {H.}~\bibnamefont {Wada}}, \ and\
		\bibinfo {author} {\bibfnamefont {Y.}~\bibnamefont {Yamada}},\ }\bibfield
	{title} {\enquote {\bibinfo {title} {$^{195}${Pt}-{NMR} evidence for opening
				of partial charge-density-wave gap in layered $\mathrm{LaPt_2Si_2}$ with
				$\mathrm{CaBe_2Ge_2}$ structure},}\ }\href {\doibase 10.7566/JPSJ.87.124713}
	{\bibfield  {journal} {\bibinfo  {journal} {J. Phys. Soc. Jpn.}\ }\textbf
		{\bibinfo {volume} {87}},\ \bibinfo {pages} {124713} (\bibinfo {year}
		{2018})}\BibitemShut {NoStop}%
	\bibitem [{\citenamefont {Das}\ \emph {et~al.}(2018)\citenamefont {Das},
		\citenamefont {Gupta}, \citenamefont {Bhattacharyya}, \citenamefont {Biswas},
		\citenamefont {Adroja},\ and\ \citenamefont {Hossain}}]{2018uSR}%
	\BibitemOpen
	\bibfield  {author} {\bibinfo {author} {\bibfnamefont {D.}~\bibnamefont
			{Das}}, \bibinfo {author} {\bibfnamefont {R.}~\bibnamefont {Gupta}}, \bibinfo
		{author} {\bibfnamefont {A.}~\bibnamefont {Bhattacharyya}}, \bibinfo {author}
		{\bibfnamefont {P.~K.}\ \bibnamefont {Biswas}}, \bibinfo {author}
		{\bibfnamefont {D.~T.}\ \bibnamefont {Adroja}}, \ and\ \bibinfo {author}
		{\bibfnamefont {Z.}~\bibnamefont {Hossain}},\ }\bibfield  {title} {\enquote
		{\bibinfo {title} {Multigap superconductivity in the charge density wave
				superconductor $\mathrm{LaPt_2Si_2}$},}\ }\href {\doibase
		10.1103/PhysRevB.97.184509} {\bibfield  {journal} {\bibinfo  {journal} {Phys.
				Rev. B}\ }\textbf {\bibinfo {volume} {97}},\ \bibinfo {pages} {184509}
		(\bibinfo {year} {2018})}\BibitemShut {NoStop}%
	\bibitem [{\citenamefont {Gupta}\ \emph {et~al.}(2019)\citenamefont {Gupta},
		\citenamefont {Thamizhavel}, \citenamefont {Rodi{\`e}re}, \citenamefont
		{Nandi}, \citenamefont {Rajeev},\ and\ \citenamefont
		{Hossain}}]{gupta2019electrical}%
	\BibitemOpen
	\bibfield  {author} {\bibinfo {author} {\bibfnamefont {R.}~\bibnamefont
			{Gupta}}, \bibinfo {author} {\bibfnamefont {A.}~\bibnamefont {Thamizhavel}},
		\bibinfo {author} {\bibfnamefont {P.}~\bibnamefont {Rodi{\`e}re}}, \bibinfo
		{author} {\bibfnamefont {S.}~\bibnamefont {Nandi}}, \bibinfo {author}
		{\bibfnamefont {K.~P.}\ \bibnamefont {Rajeev}}, \ and\ \bibinfo {author}
		{\bibfnamefont {Z.}~\bibnamefont {Hossain}},\ }\bibfield  {title} {\enquote
		{\bibinfo {title} {Electrical resistivity under pressure and thermal
				expansion of $\mathrm{LaPt_2Si_2}$ single crystal},}\ }\href {\doibase
		10.1063/1.5091784} {\bibfield  {journal} {\bibinfo  {journal} {Journal of
				Applied Physics}\ }\textbf {\bibinfo {volume} {125}},\ \bibinfo {pages}
		{143902} (\bibinfo {year} {2019})}\BibitemShut {NoStop}%
	\bibitem [{\citenamefont {Degtyareva}\ \emph {et~al.}(2007)\citenamefont
		{Degtyareva}, \citenamefont {Magnitskaya}, \citenamefont {Kohanoff},
		\citenamefont {Profeta}, \citenamefont {Scandolo}, \citenamefont {Hanfland},
		\citenamefont {McMahon},\ and\ \citenamefont {Gregoryanz}}]{Sulfur}%
	\BibitemOpen
	\bibfield  {author} {\bibinfo {author} {\bibfnamefont {O.}~\bibnamefont
			{Degtyareva}}, \bibinfo {author} {\bibfnamefont {M.~V.}\ \bibnamefont
			{Magnitskaya}}, \bibinfo {author} {\bibfnamefont {J.}~\bibnamefont
			{Kohanoff}}, \bibinfo {author} {\bibfnamefont {G.}~\bibnamefont {Profeta}},
		\bibinfo {author} {\bibfnamefont {S.}~\bibnamefont {Scandolo}}, \bibinfo
		{author} {\bibfnamefont {M.}~\bibnamefont {Hanfland}}, \bibinfo {author}
		{\bibfnamefont {M.~I.}\ \bibnamefont {McMahon}}, \ and\ \bibinfo {author}
		{\bibfnamefont {E.}~\bibnamefont {Gregoryanz}},\ }\bibfield  {title}
	{\enquote {\bibinfo {title} {Competition of charge-density waves and
				superconductivity in sulfur},}\ }\href {\doibase
		10.1103/PhysRevLett.99.155505} {\bibfield  {journal} {\bibinfo  {journal}
			{Phys. Rev. Lett.}\ }\textbf {\bibinfo {volume} {99}},\ \bibinfo {pages}
		{155505} (\bibinfo {year} {2007})}\BibitemShut {NoStop}%
	\bibitem [{\citenamefont {Zhu}\ \emph {et~al.}(2016)\citenamefont {Zhu},
		\citenamefont {Ning}, \citenamefont {Li}, \citenamefont {Ling}, \citenamefont
		{Zhang}, \citenamefont {Zhang}, \citenamefont {Wang}, \citenamefont {Liu},
		\citenamefont {Pi}, \citenamefont {Ma}, \citenamefont {Du}, \citenamefont
		{Tian}, \citenamefont {Sun}, \citenamefont {Petrovic},\ and\ \citenamefont
		{Zhang}}]{ZrTa3}%
	\BibitemOpen
	\bibfield  {author} {\bibinfo {author} {\bibfnamefont {X.~D.}\ \bibnamefont
			{Zhu}}, \bibinfo {author} {\bibfnamefont {W.}~\bibnamefont {Ning}}, \bibinfo
		{author} {\bibfnamefont {L.~J.}\ \bibnamefont {Li}}, \bibinfo {author}
		{\bibfnamefont {L.~S.}\ \bibnamefont {Ling}}, \bibinfo {author}
		{\bibfnamefont {R.~R.}\ \bibnamefont {Zhang}}, \bibinfo {author}
		{\bibfnamefont {J.~L.}\ \bibnamefont {Zhang}}, \bibinfo {author}
		{\bibfnamefont {K.~F.}\ \bibnamefont {Wang}}, \bibinfo {author}
		{\bibfnamefont {Y.}~\bibnamefont {Liu}}, \bibinfo {author} {\bibfnamefont
			{L.}~\bibnamefont {Pi}}, \bibinfo {author} {\bibfnamefont {Y.~C.}\
			\bibnamefont {Ma}}, \bibinfo {author} {\bibfnamefont {H.~F.}\ \bibnamefont
			{Du}}, \bibinfo {author} {\bibfnamefont {M.~L.}\ \bibnamefont {Tian}},
		\bibinfo {author} {\bibfnamefont {Y.~P.}\ \bibnamefont {Sun}}, \bibinfo
		{author} {\bibfnamefont {C.}~\bibnamefont {Petrovic}}, \ and\ \bibinfo
		{author} {\bibfnamefont {Y.~H.}\ \bibnamefont {Zhang}},\ }\bibfield  {title}
	{\enquote {\bibinfo {title} {Superconductivity and charge density wave in
				$\mathrm{ZrTe_{3-x}Se_3}$},}\ }\href {\doibase 10.1038/srep26974} {\bibfield
		{journal} {\bibinfo  {journal} {Scientific Reports}\ }\textbf {\bibinfo
			{volume} {6}},\ \bibinfo {pages} {26974} (\bibinfo {year}
		{2016})}\BibitemShut {NoStop}%
	\bibitem [{\citenamefont {Rice}(1968)}]{A_FL}%
	\BibitemOpen
	\bibfield  {author} {\bibinfo {author} {\bibfnamefont {M.~J.}\ \bibnamefont
			{Rice}},\ }\bibfield  {title} {\enquote {\bibinfo {title} {Electron-electron
				scattering in transition metals},}\ }\href {\doibase
		10.1103/PhysRevLett.20.1439} {\bibfield  {journal} {\bibinfo  {journal}
			{Phys. Rev. Lett.}\ }\textbf {\bibinfo {volume} {20}},\ \bibinfo {pages}
		{1439--1441} (\bibinfo {year} {1968})}\BibitemShut {NoStop}%
	\bibitem [{\citenamefont {Regueiro}\ \emph {et~al.}(1992)\citenamefont
		{Regueiro}, \citenamefont {Mignot},\ and\ \citenamefont
		{Castello}}]{1992EPL}%
	\BibitemOpen
	\bibfield  {author} {\bibinfo {author} {\bibfnamefont {M.~N{\'u}nez}\
			\bibnamefont {Regueiro}}, \bibinfo {author} {\bibfnamefont {J.~M.}\
			\bibnamefont {Mignot}}, \ and\ \bibinfo {author} {\bibfnamefont
			{D.}~\bibnamefont {Castello}},\ }\bibfield  {title} {\enquote {\bibinfo
			{title} {Superconductivity at high pressure in $\mathrm{NbSe_3}$},}\ }\href
	{\doibase 10.1209/0295-5075/18/1/010} {\bibfield  {journal} {\bibinfo
			{journal} {EPL (Europhysics Letters)}\ }\textbf {\bibinfo {volume} {18}},\
		\bibinfo {pages} {53} (\bibinfo {year} {1992})}\BibitemShut {NoStop}%
	\bibitem [{\citenamefont {Hamlin}(2015)}]{hamlinSC}%
	\BibitemOpen
	\bibfield  {author} {\bibinfo {author} {\bibfnamefont {J.~J.}\ \bibnamefont
			{Hamlin}},\ }\bibfield  {title} {\enquote {\bibinfo {title}
			{Superconductivity in the metallic elements at high pressures},}\ }\href
	{\doibase 10.1016/j.physc.2015.02.032} {\bibfield  {journal} {\bibinfo
			{journal} {Physica C: Superconductivity and its Applications}\ }\textbf
		{\bibinfo {volume} {514}},\ \bibinfo {pages} {59--76} (\bibinfo {year}
		{2015})}\BibitemShut {NoStop}%
	\bibitem [{\citenamefont {Monteverde}\ \emph {et~al.}(2001)\citenamefont
		{Monteverde}, \citenamefont {Nunez-Regueiro}, \citenamefont {Rogado},
		\citenamefont {Regan}, \citenamefont {Hayward}, \citenamefont {He},
		\citenamefont {Loureiro},\ and\ \citenamefont
		{Cava}}]{monteverde2001pressure}%
	\BibitemOpen
	\bibfield  {author} {\bibinfo {author} {\bibfnamefont {M.}~\bibnamefont
			{Monteverde}}, \bibinfo {author} {\bibfnamefont {M.}~\bibnamefont
			{Nunez-Regueiro}}, \bibinfo {author} {\bibfnamefont {N.}~\bibnamefont
			{Rogado}}, \bibinfo {author} {\bibfnamefont {K.~A.}\ \bibnamefont {Regan}},
		\bibinfo {author} {\bibfnamefont {M.~A.}\ \bibnamefont {Hayward}}, \bibinfo
		{author} {\bibfnamefont {T.}~\bibnamefont {He}}, \bibinfo {author}
		{\bibfnamefont {S.~M.}\ \bibnamefont {Loureiro}}, \ and\ \bibinfo {author}
		{\bibfnamefont {R.~J.}\ \bibnamefont {Cava}},\ }\bibfield  {title} {\enquote
		{\bibinfo {title} {Pressure dependence of the superconducting transition
				temperature of magnesium diboride},}\ }\href {\doibase
		10.1126/science.1059775} {\bibfield  {journal} {\bibinfo  {journal}
			{Science}\ }\textbf {\bibinfo {volume} {292}},\ \bibinfo {pages} {75--77}
		(\bibinfo {year} {2001})}\BibitemShut {NoStop}%
	\bibitem [{\citenamefont {Hull}\ \emph {et~al.}(1981)\citenamefont {Hull},
		\citenamefont {Wernick}, \citenamefont {Geballe}, \citenamefont {Waszczak},\
		and\ \citenamefont {Bernardini}}]{LPG_SC}%
	\BibitemOpen
	\bibfield  {author} {\bibinfo {author} {\bibfnamefont {G.~W.}\ \bibnamefont
			{Hull}}, \bibinfo {author} {\bibfnamefont {J.~H.}\ \bibnamefont {Wernick}},
		\bibinfo {author} {\bibfnamefont {T.~H.}\ \bibnamefont {Geballe}}, \bibinfo
		{author} {\bibfnamefont {J.~V.}\ \bibnamefont {Waszczak}}, \ and\ \bibinfo
		{author} {\bibfnamefont {J.~E.}\ \bibnamefont {Bernardini}},\ }\bibfield
	{title} {\enquote {\bibinfo {title} {Superconductivity in the ternary
				intermetallics $\mathrm{YbPd_2Ge_2}$, $\mathrm{LaPd_2Ge_2}$, and
				$\mathrm{LaPt_2Ge_2}$},}\ }\href {\doibase 10.1103/PhysRevB.24.6715}
	{\bibfield  {journal} {\bibinfo  {journal} {Phys. Rev. B}\ }\textbf {\bibinfo
			{volume} {24}},\ \bibinfo {pages} {6715--6718} (\bibinfo {year}
		{1981})}\BibitemShut {NoStop}%
	\bibitem [{\citenamefont {Maeda}\ \emph {et~al.}(2015)\citenamefont {Maeda},
		\citenamefont {Matano}, \citenamefont {Yatagai}, \citenamefont {Oguchi},\
		and\ \citenamefont {Zheng}}]{PRBLaPt2Ge2}%
	\BibitemOpen
	\bibfield  {author} {\bibinfo {author} {\bibfnamefont {S.}~\bibnamefont
			{Maeda}}, \bibinfo {author} {\bibfnamefont {K.}~\bibnamefont {Matano}},
		\bibinfo {author} {\bibfnamefont {R.}~\bibnamefont {Yatagai}}, \bibinfo
		{author} {\bibfnamefont {T.}~\bibnamefont {Oguchi}}, \ and\ \bibinfo {author}
		{\bibfnamefont {G.~Q.}\ \bibnamefont {Zheng}},\ }\bibfield  {title} {\enquote
		{\bibinfo {title} {Superconductivity and the electronic phase diagram of
				$\mathrm{LaPt_{2-x}Ge_{2+x}}$},}\ }\href {\doibase
		10.1103/PhysRevB.91.174516} {\bibfield  {journal} {\bibinfo  {journal} {Phys.
				Rev. B}\ }\textbf {\bibinfo {volume} {91}},\ \bibinfo {pages} {174516}
		(\bibinfo {year} {2015})}\BibitemShut {NoStop}%
\end{thebibliography}
\end{document}